\documentclass[10pt,letterpaper]{article}
\usepackage[top=0.85in,left=2.75in,footskip=0.75in]{geometry}

\usepackage{amsmath,amssymb}

\usepackage{changepage}

\usepackage[utf8x]{inputenc}

\usepackage{textcomp,marvosym}

\usepackage{cite}

\usepackage{nameref,hyperref}

\usepackage[right]{lineno}

\usepackage{microtype}
\DisableLigatures[f]{encoding = *, family = * }

\usepackage[table]{xcolor}

\usepackage{array}

\newcolumntype{+}{!{\vrule width 2pt}}

\newlength\savedwidth

\raggedright
\usepackage[aboveskip=1pt,labelfont=bf,labelsep=period,justification=raggedright,singlelinecheck=off]{caption}

\makeatletter
\renewcommand{\@biblabel}[1]{\quad#1.}
\makeatother

\date{}

\usepackage{lastpage,fancyhdr,graphicx}
\usepackage{epstopdf}
\fancyheadoffset[L]{2.25in}
\fancyfootoffset[L]{2.25in}
\begin{document}
\vspace*{0.2in}

\begin{flushleft}
{\Large
\textbf\newline{Community Detection in Dynamic Networks via Adaptive Label Propagation} 
}
\newline
\\
Jihui Han\textsuperscript{1,2*},
Wei Li\textsuperscript{1*},
Longfeng Zhao\textsuperscript{1},
Zhu Su\textsuperscript{1},
Yijiang Zou\textsuperscript{1},
Weibing Deng\textsuperscript{1*}
\\
\bigskip
\textbf{1} Complexity Science Center \& Institute of Particle Physics, Central China Normal University, Wuhan, Hubei, China
\\
\textbf{2} School of Computer and Communication Engineering, Zhengzhou University of Light Industry, Zhengzhou, Henan, China
\\
\bigskip

%
%





* Corresponding author \\
jh@mails.ccnu.edu.cn (JH), liw@mail.ccnu.edu.cn (WL), wdeng@mail.ccnu.edu.cn (WD)

\end{flushleft}
\section*{Abstract}
An adaptive label propagation algorithm (ALPA) is proposed to detect and monitor communities in dynamic networks. Unlike the traditional methods by re-computing the whole community decomposition after each modification of the network, ALPA takes into account the information of historical communities and updates its solution according to the network modifications via a local label propagation process, which generally affects only a small portion of the network. This makes it respond to network changes at low computational cost. The effectiveness of ALPA has been tested on both synthetic and real-world networks, which shows that it can successfully identify and track dynamic communities. Moreover, ALPA could detect communities with high quality and accuracy compared to other methods. Therefore, being low-complexity and parameter-free, ALPA is a scalable and promising solution for some real-world applications of community detection in dynamic networks.



\section*{Introduction}
Many real-world systems can be represented as networks \cite{Strogatz2001,RevModPhys.74.47,doi:10.1080/00018730110112519,doi:10.1137/S003614450342480}, in which nodes represent individuals and edges represent the relationships or interactions between individuals, such as the Internet \cite{Faloutsos:1999:PRI:316194.316229}, friendship networks \cite{Amaral10102000}, collaboration networks \cite{Newman16012001}, food webs \cite{Williams2000,Dunne01102002}, and metabolic networks \cite{Jeong2000,Fell2000}.

Community structure is a prominent feature of networks and has received much attention in recent years. It deepens our understanding of the underlying structure of many real-world networks \cite{Faloutsos:1999:PRI:316194.316229,Amaral10102000,Newman16012001,Williams2000,Dunne01102002}, and promises a variety of practical applications ranging from the determination of functional modules within neural networks to the analysis of communities on the Internet. A network is deemed to have community structure if it can be easily divided into groups of nodes with denser connections internally and sparser connections between groups \cite{Girvan11062002,1742-5468-2005-09-P09008,Fortunato201075}. Detecting community structure is a challenging task and many algorithms have been developed in the last decade, such as modularity optimization \cite{Newman06062006,FastUnfolding2008}, dynamic label propagation \cite{PhysRevE.76.036106,1367-2630-12-10-103018,JieruiXie-LabelRank-NSW:2013}, statistical inference \cite{PhysRevE.83.016107,Aldecoa2013surprise,10.1371/journal.pone.0018961}, spectral clustering \cite{Singh2015}, information-theoretic methods \cite{Rosvall01052007,Rosvall29012008}, and topology based \cite{Zalik2015,PhysRevE.72.046108} methods.

However, most of the methods treat the network as a static one which is derived from aggregating data during a long period of time. In this way, the evolutionary information of the network and its communities is lost because real-world networks are always evolving, either by adding or removing nodes or edges over time. These static methods cannot tell us how communities evolve over time by neglecting intrinsic evolution of the network. Moreover, if one would like to monitor the communities of a network in real time, the static methods are commonly time-consuming as they have to compute the whole community decomposition even if a very small modification of the network occurs, especially when the network evolves rapidly.

One way to analyze communities in a dynamic or evolving network is to slice the network into many snapshots, whichever is a static network. Algorithms along this line first analyze snapshots of the dynamic network at different time steps more or less independently, and then compare communities of different snapshots with each other so that one can monitor the evolution of each community \cite{5562773,Asur:2007:EFC:1281192.1281290,ning2007incremental,Palla2007,6758385,Lin:2009:ACE:1514888.1514891}. For example, one of such algorithms, FacetNet \cite{Lin:2009:ACE:1514888.1514891} detects dynamic communities by optimizing a quality function which considers both the quality and the stability of communities. While another one, DSBM \cite{yang2009bayesian} fits the evolving network to a dynamic version of the stochastic block model, and determines the community assignment by estimating the parameters of the model. A main disadvantage of these algorithms is that they are commonly time-consuming when the network evolves rapidly and the time slices are extremely small, i.e., the network has a lot of snapshots to be computed. Moreover, it is difficult to find the appropriate time window for dividing the dynamic network into static snapshots.

Another way is to adaptively update the current community structure based on previous ones according to modifications of the network. These algorithms quickly adapt their results when the network undergoes a slight modification rather than compute the whole community decomposition from scratch \cite{Sun:2007:GPM:1281192.1281266,iLCD,10.1371/journal.pone.0091431,10.1371/journal.pone.0086891}. For example, Nguyen NP et al. \cite{10.1371/journal.pone.0091431} proposed a modularity-based algorithm named \emph{quick community adaptive} (QCA) which greedily changes memberships of nodes by optimizing a local modularity function whenever a small modification occurs in the network. A similar kind of algorithm LabelRankT \cite{xie2013labelrankt} adjusts its detecting results according to the network modifications through a stabilized label propagation process by taking advantage of what is already obtained in previous snapshots. Another algorithm iLCD \cite{iLCD} first determines whether or not the new node joins existing communities according to two adaptive threshold conditions, then decides whether a new edge is able to form a minimal community or not, and finally merges all communities that are very close to each other (i.e. they have more than a certain ratio of common nodes). If updates are computed efficiently, these adaptive methods are commonly more efficient than computing communities on each snapshot separately when used to monitor large dynamic networks in real time.

Although these methods have been developed to analyze communities in dynamic networks, most of them are not applicable to analyzing real-world networks because they either need to know the prior information of communities (e.g., the number of them) which is usually unknown in advance, or require some user-defined parameters which are difficult to be set in practice.

The present paper proposes an adaptive label propagation algorithm (ALPA) for analyzing dynamic communities with no need for the prior information of communities or user-defined parameters. It detects communities by taking into account their evolutions, and updates the current community structure through a local label propagation process, which only affects a small portion of the network. Therefore, ALPA can efficiently respond to network modifications at low computational cost. Moreover, ALPA is an incremental algorithm, and naturally works in a streaming manner. We evaluated the proposed method on both synthetic and real-world networks. Experimental results show that our method detects communities with high quality and successfully tracks their evolution over time. ALPA has been implemented in a freely available Julia package released under the MIT License (\url{https://github.com/afternone/ALPA.jl}), and we believe it would be a helpful tool in the analysis of dynamic networks.

The rest of this paper is organized as follows. The next section depicts the details of the proposed method. Then, the method is tested on both synthetic and read-world networks. Finally, we summarize our findings in the last section.

\section*{Methods}
\subsection*{Local label propagation (LLP)}

The LLP process uses LPA’s \cite{PhysRevE.76.036106} label propagation technique to propagate labels only throughout part of the network. It maintains an active node list that contains all currently active nodes and finishes execution when the list is empty. An active node is the one whose label is not the majority one among its neighbors and potentially changes its label if it was to attempt an update. The LLP process asynchronously updates each label of the nodes in the active node list in a random order according to the generalized update rule proposed by Xie and Szymanski \cite{6004645}, in which the positive neighborhood strength is taken into account when a node considers a new label. During the process, if a node changes its label after an update (i.e., the node is still active), all its neighbors will be inserted into the active node list. If the node turns inactive (i.e., it does not change its label after an update), it will then be removed from the active node list. As this process goes on, the active node list will eventually become empty.



\begin{figure}[!h]
	\centering
	\includegraphics[width=\linewidth]{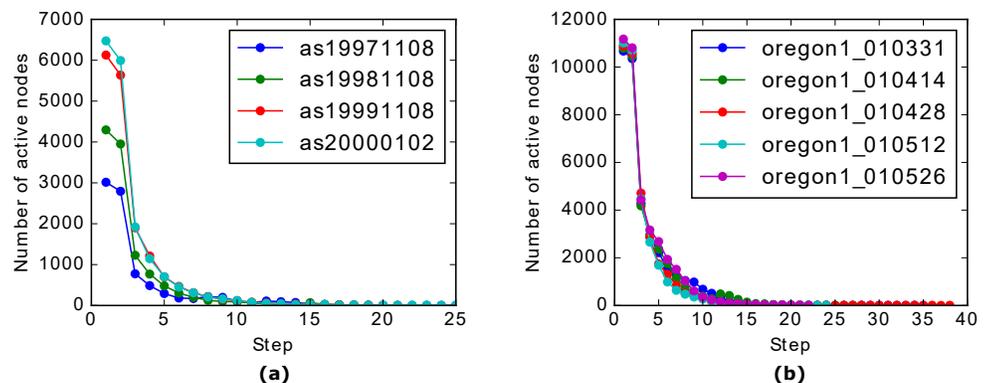}
	\caption{(Color online) {\bf Convergence behavior of the LLP process on real-world networks.} 
	(a) and (b) show the number of active nodes per step during the LLP process on some snapshots of AS-Internet and AS-Oregon datasets, respectively. The labels in the legend indicate the date of the snapshots.}
	\label{fig1}
\end{figure}

To analyze the convergence behavior of the LLP process, we perform it on some snapshots of AS-Internet and AS-Oregon datasets \cite{Leskovec:2005:GOT:1081870.1081893}. For simplicity, we assume that all nodes are active at the beginning. During the LLP process, we record the number of active nodes per step and show the convergence history in Fig \ref{fig1}. As one can see, the LLP process converges quickly. The number of active nodes decreases dramatically in the first few steps.

\subsection*{Updating existing communities according to network modifications}
\begin{figure}[!h]
	\centering
	\includegraphics[width=\linewidth]{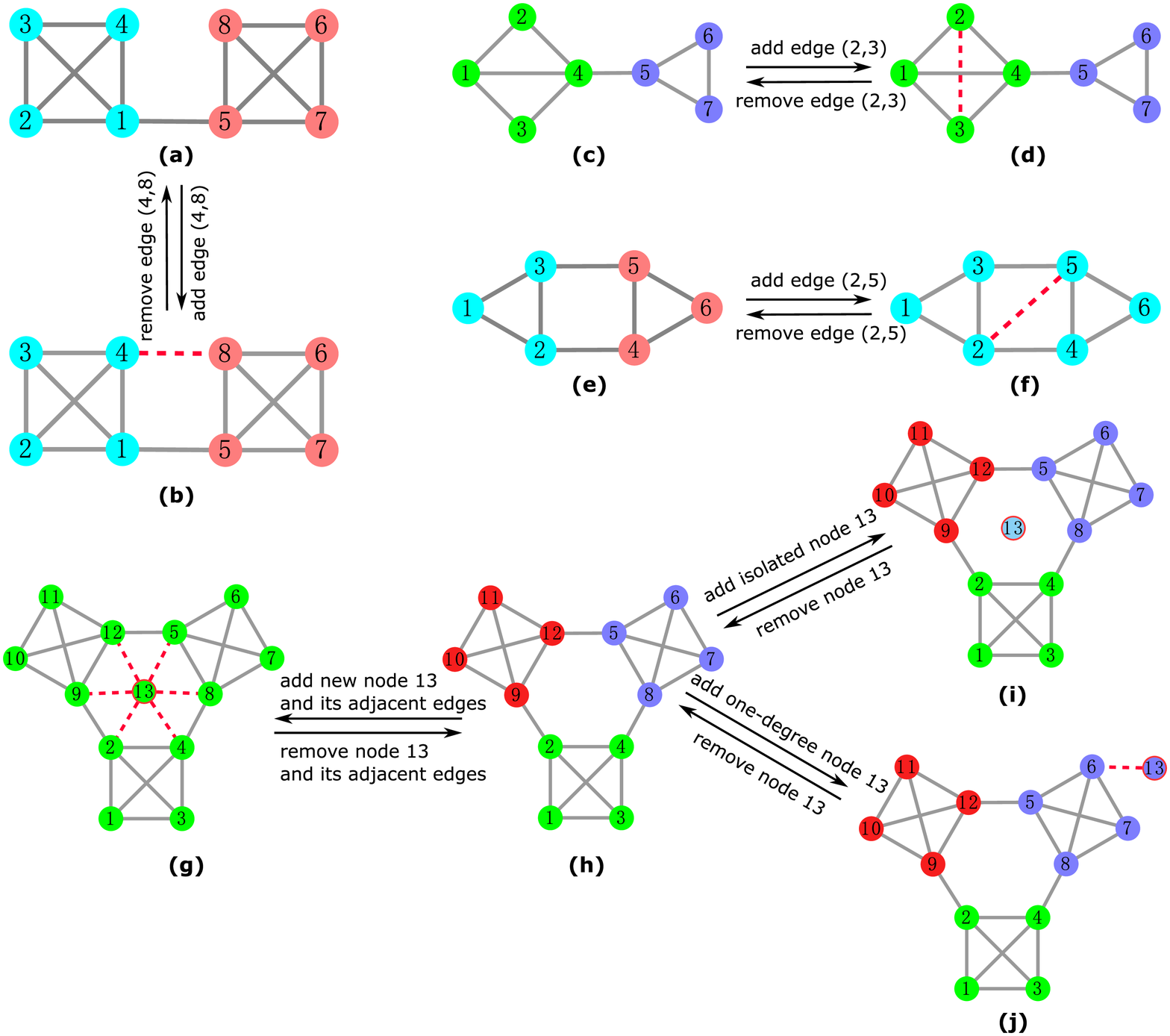}
	\caption{(Color online) {\bf Schematic illustrations of how community structure can be affected by network modifications.} 
	The node to be added or removed is stroked by red, and the edge to be added or removed is represented by red dashed line. Communities are distinguished by different filled colors of the nodes.}
	\label{fig2}
\end{figure}

We show the details of the procedures concerning different modifications of the network. When a new edge connecting two existing nodes is added, there are two cases: intra-community edge or inter-community edge. An intra-community edge’s addition will tighten up the community and should not change the current partition (see Fig \ref{fig2} (c\textrightarrow d)). However, an inter-community edge’s addition could potentially move one of the two endpoints from the current community into another, or merge the two communities into a new and large one (see Fig \ref{fig2} (e\textrightarrow f) for an example). To handle this, we first relabel all nodes in the two corresponding communities which are connected by the new edge. For convenience, we just relabel them with their IDs. Then we insert all nodes of the two target communities into the active node list. Finally, we apply LLP process to update the community structure. In order to avoid unnecessary updates, after adding a new edge, if each node of the two endpoints still has more connections within its community than its connections with other nodes which do not belong to its commnunity, we will not carry out the LLP process.

\begin{figure}[!h]
	\centering
	\includegraphics[width=0.6\linewidth]{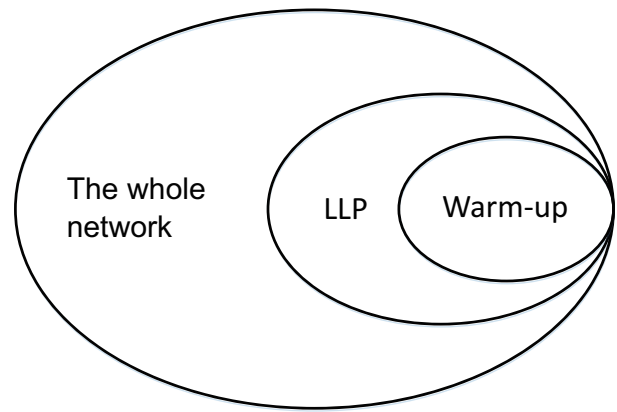}
	\caption{{\bf Schematic diagram of the scope of the warm-up step and the LLP process.} 
	The warm-up step only propagates labels inside the target community (or communities), while the LLP process may involve some nodes outside of the target community (or communities). These two processes generally affect a small portion of the whole network, especially when the network has many small communities.}
	\label{fig3}
\end{figure}

However, these operations alone are still not enough to guarantee good performance in practice. This is because when nodes in the border of these two communities update their labels at the initial stage, they tend to adopt more frequent labels from other adjacent communities. Consequently, communities tend to merge during the evolution of the network. To deal with this issue, prior to the LLP process, we perform a ``warm-up" step in which the target communities are treated as a subgraph for labels to propagate, i.e., during the label propagation process, when we update the label of a node, we only consider labels of its neighbors that are in the subgraph. Note that nodes in the target community (communities) need to reinitialize their labels before the warm-up step, and the LLP process is then based on the labels obtained in the warm-up step. This strategy allows us to preserve detected communities. Obviously, the number of affected nodes of warm-up step should not be larger than that of the LLP process. There is generally a certain relationship among the warm-up step, the LLP process, and the whole network, as shown in  Fig \ref{fig3}.

When an existing edge is deleted, there are also two cases: the edge is either inter-community or intra-community. For the former case, the deletion will make the current community structure clearer (see Fig \ref{fig2} (b\textrightarrow a) for an example). Herewith we leave the partition intact. For the latter case, the deletion may break the community into small pieces (see Fig \ref{fig2} (f\textrightarrow e) for an example), and these pieces could join in other communities. To deal with this, we first insert all nodes of the target community into the active node list, and then perform the warm-up and LLP processes to update the community structure.

If an isolate node is inserted, we simply create a new community for it (see Fig \ref{fig2} (h\textrightarrow i)). While, if the node comes with some adjacent edges connecting to one or more existing communities (see Fig \ref{fig2} (h\textrightarrow g)), we split the process into two steps, i.e., first add an ``isolate" node and then add its adjacent edges one after another. If an isolated node is removed, the current community structure will be unchanged (see Fig \ref{fig2} (i\textrightarrow h)). However, when a node with degree larger than or equal to two is removed (see Fig \ref{fig2} (g\textrightarrow h)), all its adjacent edges will be destroyed, and the community containing the node could remain unchanged, or break into a number of small pieces which could join in other communities. To efficiently deal with this case, we first remove all the node’s adjacent edges one by one, and then remove the node itself.

Finally, combining all these cases, our adaptive label propagation algorithm is summarized as follows.
\begin{enumerate}
	\item\label{step1} Initialize an empty graph and an empty partition.
	\item\label{step2} For each modification:
	\begin{enumerate}
		\item\label{step2a} If a new edge is added, we update the current partition according to the procedure of adding a new edge.
		\item\label{step2b} If an existing edge is removed from the network, we update the current partition according to the procedure of removing an existing edge.
		\item\label{step2c} If an isolated node is added, we simply create a new community for it.
		\item\label{step2d} If a node with some associate edges is added, we first create a new community for it (i.e., step \ref{step2c}), and then add all its associate edges one by one according to step \ref{step2a}.
		\item\label{step2e} If an isolated node is removed, we just delete it from the current partition, and leave other communities intact.
		\item\label{step2f} If a node with degree larger than one is removed, we first remove all its adjacent edges one by one according to step \ref{step2b}, and then remove the node itself according to step \ref{step2e}.
	\end{enumerate}
	\item\label{step3} Output the graph and its partition at each time step.
\end{enumerate}	

Note that, ALPA can also start with any given snapshot network instead of an empty one. The initial community structure can be obtained with any of the available static methods, or with ALPA itself (i.e., start with an empty graph and treat the network as a collection of new nodes or edges).

\section*{Results}
In this section, we first evaluate our method on different synthetic networks with known community structures, and then show the results on two popular real-world datasets: AS-Internet and AS-Oregon. In order to verify the performance of our method, we compare it with two public available methods FacetNet \cite{Lin:2009:ACE:1514888.1514891} and iLCD \cite{iLCD}. Moreover, a widely used static method Infomap \cite{Rosvall29012008} is also involved in the comparisons.

\subsection*{Synthetic networks}
\begin{figure}[!h]
	\centering
	\includegraphics[width=\linewidth]{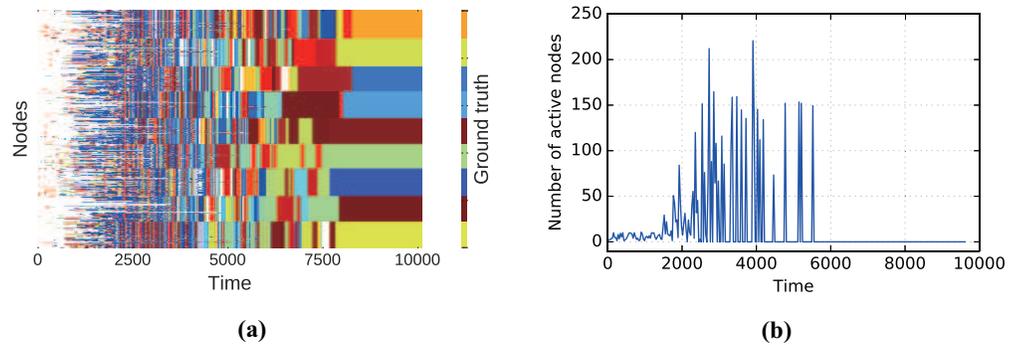}
	\caption{(Color online) {\bf Incremental detection of communities on the LFR network.} 
	The color code in panel (a) corresponds to our incremental detection of the different communities in the network over time. Each node (vertical axes) at every time step belongs to a community, which is distinguished by different colors. White indicates that the node does not exist at that snapshot. From left to right, as new edges are added, nodes tends to group together, hence, many colors disappear and the community structure becomes clear. Eventually, our algorithm could correctly detect the true community structure. In particular, the number of active nodes involved in each event is shown in panel (b). The parameters of the LFR network are: $N=1000$, $\mu=0.3$, $\langle k\rangle=20$, $k_{max}=50$, $\gamma=2$, $\beta=1$, minimum and maximum community sizes are 90 and 125 respectively.}
	\label{fig4}
\end{figure}

We start with the first synthetic network which is a static network generated by the well-known Lancichinetti-Fortunato-Radicchi (LFR) benchmark \cite{PhysRevE.78.046110}, to show that our algorithm can handle incremental inputs. The network contains 1000 nodes which are naturally grouped into nine communities. There are around 10000 edges in the network. Starting with an empty network, we add these edges one by one and use our algorithm to update the community structure after each edge’s addition. Fig \ref{fig4} (a) shows the evolution of communities in the network. It is shown that ALPA identifies the nine communities correctly. In addition, we are interested in knowing how many nodes are involved in ALPA after each edge’s addition, which can be used to estimate the time complexity of our algorithm. We record the number of involved nodes (i.e., those are activated at least once) in each time step and plot them in Fig \ref{fig4} (b). As one can see, most of the modifications only affect a few nodes. The average number of involved nodes in each time step is 23.7, which is tiny compared to the network size, so our algorithm can efficiently respond to the changes in network topology.

\begin{figure}[!h]
	\centering
	\includegraphics[width=\linewidth]{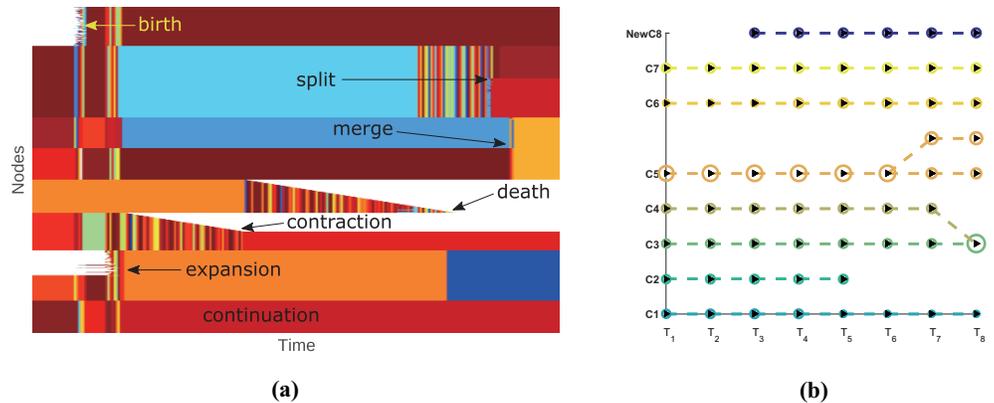}
	\caption{(Color online) {\bf Visualization of communities' evolution for time-varying LFR network.} 
	The color code in panel (a) corresponds to our adaptive detection of communities in the network over time. Each node (vertical axes) at every time step belongs to a community, which is distinguished by different colors. White indicates that the node does not exist at that snapshot. From left to right, embedded community events occur sequentially. Panel (b) visualizes the communities' evolution by using Netgram with parameters $\rho=0.4$ and $\nu=0.1$ \cite{10.1371/journal.pone.0137502}. The parameters of the LFR network are the same as those in Fig \ref{fig4}.}
	\label{fig5}
\end{figure}

The second synthetic network is also an LFR network, but with embedded community events inside it. The network is constructed according to the following steps. Starting with a static LFR network, at each time step, we make a slight modification to the network, such as nodes or edges’ addition or removal. In this way we produce a dynamic network which contains some community events. For the experiment conducted here, community events are sequentially embedded as follows: birth, expansion, shrinkage, death, separation and combination. As shown in Fig \ref{fig5} (a), our algorithm recognizes all these major changes of communities and successfully tracks the evolution of each community. To demonstrate the evolution of each community clearer, we select eight snapshots of the network and visualize them using Netgram tool developed by Mall R et al \cite{10.1371/journal.pone.0137502}. As shown in Fig \ref{fig5} (b), each circle represents a community, and its size is proportional to the number of nodes inside the community at that time step. The dashed line represents the evolution of communities between two consecutive time steps. We can see that a new community \emph{NewC8} is born at snapshot $T_3$, \emph{C6} expands at $T_4$, \emph{C2} disappears at $T_5$, \emph{C1} shrinks at $T_4$, \emph{C5} is divided into two small communities at $T_6$, \emph{C3} and \emph{C4} are merged at $T_8$, while community \emph{C7} remains intact throughout the whole evolution.

In order to compare ALPA with other methods, we employ the dynamic benchmark model proposed by Granell et al \cite{PhysRevE.92.012805} to generate three standard benchmarks: grow-shrink, merge-split and mixed. The first one contains communities that grow and shrink periodically in size, while the second one considers communities that merge and split periodically. The third one is a mixed version of the previous two and consists of a combination of all the four operations. Each of the benchmark network consists of 100 time steps, and is divided into 4 communities, where each community has 32 nodes initially (therefore the total size of the network is 128). The nodes of the same community are connected with a probability $p_{in}=0.05$, whereas nodes of different communities are connected with a probability $p_{out}=0.5$. For the grow-shrink process, the maximum fraction of nodes moving from one community to another is 0.5, i.e., there are at most 16 nodes switching between communities.

\begin{figure}[!h]
	\centering
	\includegraphics[width=\linewidth]{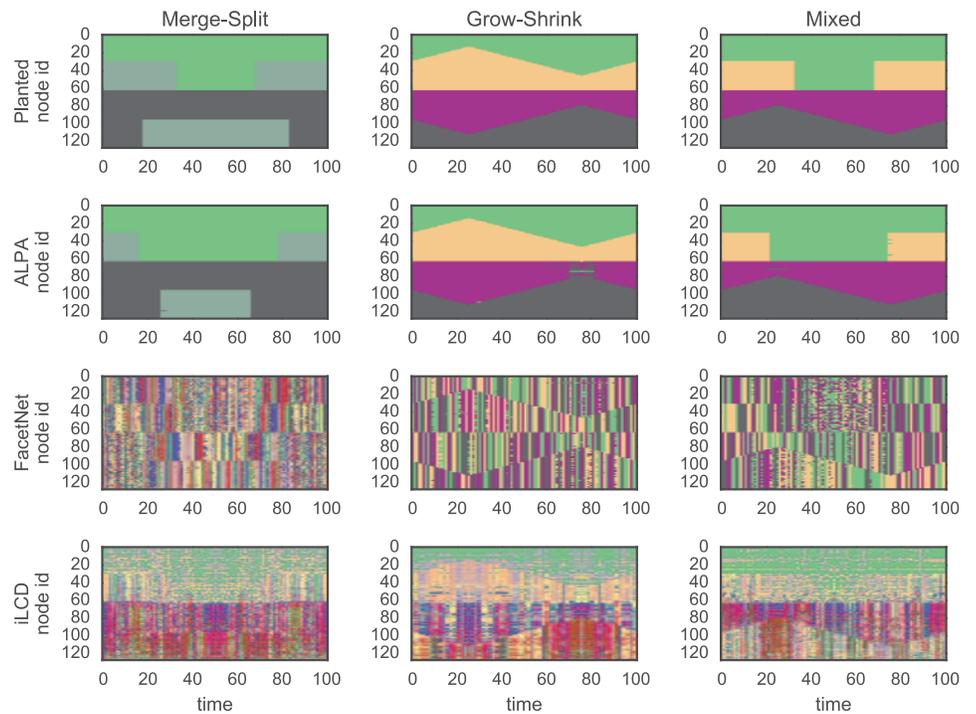}
	\caption{(Color online) {\bf Results of the application of different methods on the three standard benchmarks (in columns).} The first row corresponds to the planted partition of each benchmark, while the three remaining rows are the partitions detected by different algorithms. In each plot, the vertical axis corresponds to the index of nodes in the network, while the horizontal axis represents the time. The color of each pair \{node, time\} indicates the community to which the node is belongs at that specific time.}
	\label{fig6}
\end{figure}

Fig \ref{fig6} shows the planted partitions and the results from different algorithms. It can be seen that the results of ALPA are mostly correct, except for some extreme time steps, whereas the partitions detected by FacetNet and iLCD are very different from the planted ones. Moreover, the partitions detected by ALPA have higher consistency through time than those detected by the other two algorithms.

\begin{figure}[!h]
	\centering
	\includegraphics[width=\linewidth]{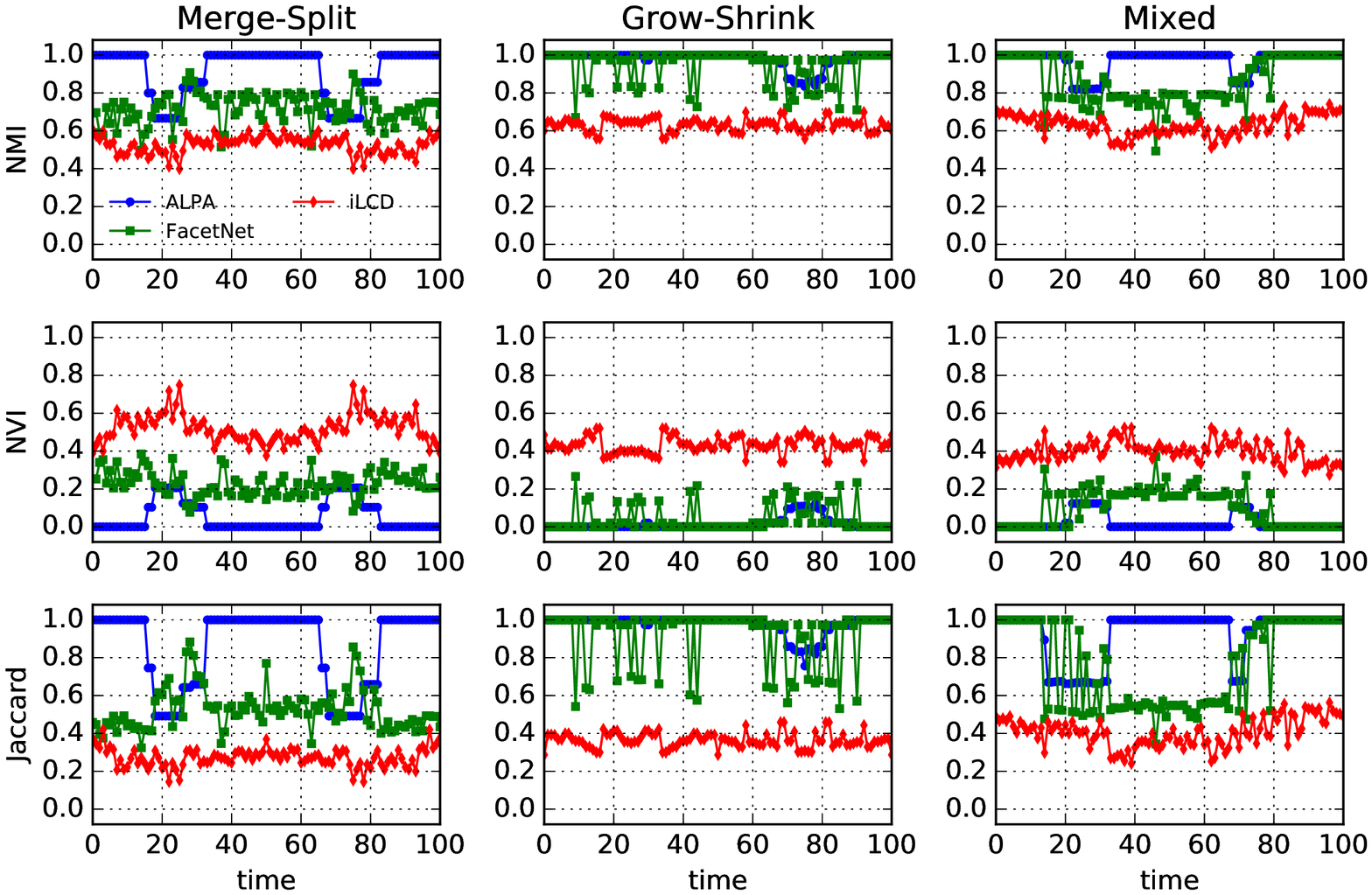}
	\caption{(Color online) {\bf Three different measures (NMI, NVI and Jaccard index) between the planted partitions and the partitions detected by different algorithms for the three standard benchmarks.} There is a column for each benchmark and a row for each measure.}
	\label{fig7}
\end{figure}

To quantitatively evaluate the results, we calculate the normalized mutual information (NMI) \cite{1742-5468-2005-09-P09008}, the normalized variation of information (NVI) and the Jaccard index between the planted partitions and the detected ones. As shown in Fig \ref{fig7}, in most snapshots, the values of NMI and Jaccard index of ALPA are higher than those of the other two algorithms, while the values of NVI of ALPA are lower than those of the other two algorithms. These results indicate that ALPA outperforms FacetNet and iLCD on these dynamic benchmark networks.

\begin{figure}[!h]
	\centering
	\includegraphics[width=\linewidth]{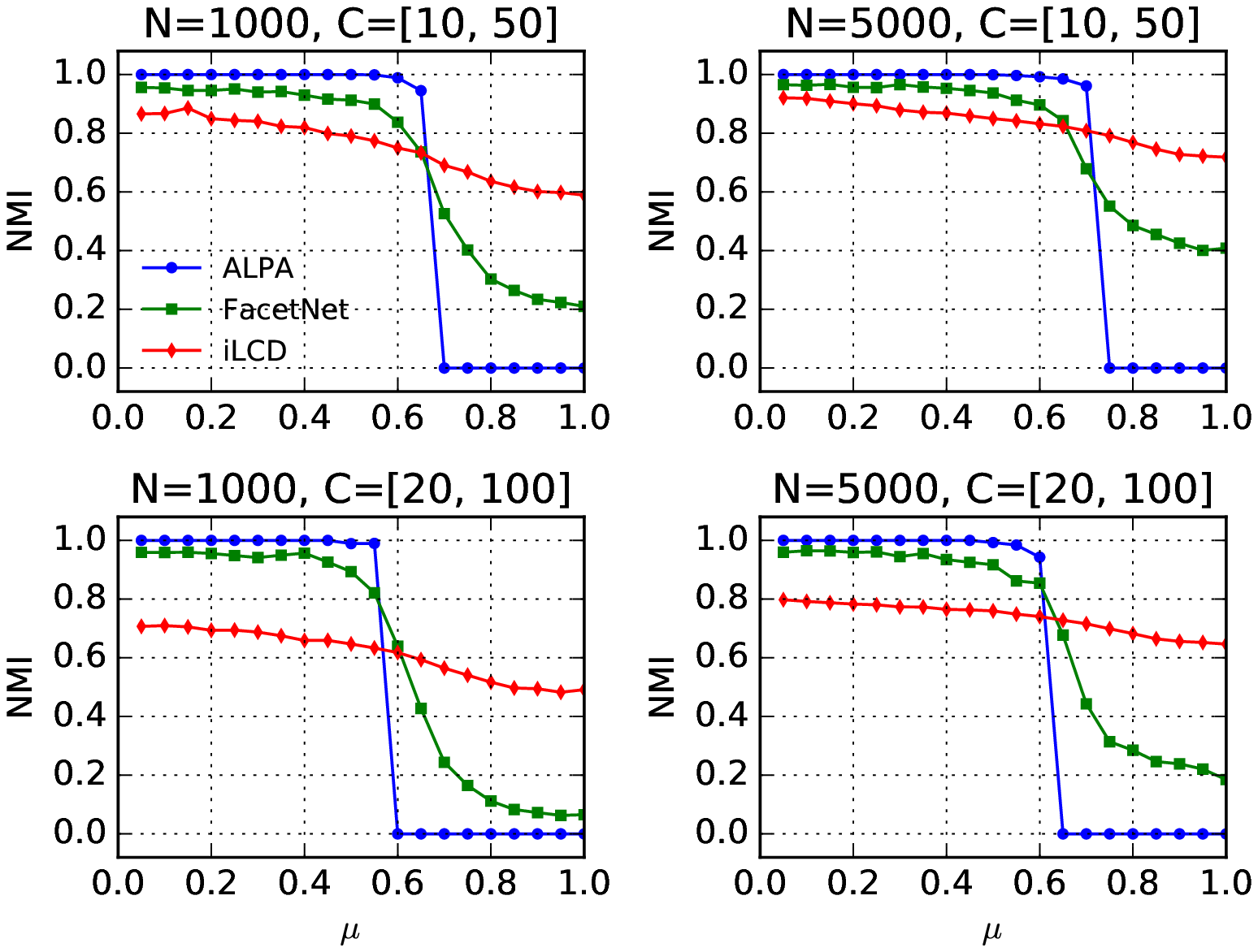}
	\caption{(Color online) {\bf Normalized mutual information (NMI) as a function of the mixing parameter $\mu$ in LFR networks.} Four network scenarios are shown, which correspond to two different network sizes ($N=1000,5000$) and, for a given size, to two different ranges for the community sizes ($C=[10,50],[20,100]$). Each point on the curves corresponds to the average value of the NMI value over 100 network realizations.}
	\label{fig8}
\end{figure}

In order to compare the performance of different algorithms for community detection on general settings, we test them on the LFR networks with four scenarios: two different network sizes (1000 and 5000 nodes) and two different ranges of community sizes ([10,50] and [20,100]). The following parameters are the same for all the LFR networks used here: the average and maximum degrees are 20 and 50 respectively, the power-law exponent of the degree distribution and the community size distribution are -2 and -1 respectively, and the mixing parameter increases from 0 to 1 with step size being 0.05. For iLCD, if a node belongs to multiple communities, we assign it to the one with maximum size to output disjoint partition. Since FacetNet requires the number of communities as its input parameter, we assign its value with the number of planted communities. We use the NMI to measure the consistency between the planted partition and the detected partition. It can be found (see Fig \ref{fig8}) that ALPA detects communities correctly and outperforms the other two algorithms up to $\mu\sim 0.6$ in all cases.


\subsection*{Real-world networks}
In this section, we tested the performance of ALPA on two real-world networks from the Stanford network analysis project datasets \cite{snapnets}. AS-Internet \cite{Leskovec:2005:GOT:1081870.1081893} and AS-Oregon \cite{Leskovec:2005:GOT:1081870.1081893} were chosen from the available datasets, since they have a varying number of snapshots. The description of the two datasets is as follows.

\begin{figure}[!h]
	\centering
	\includegraphics[width=\linewidth]{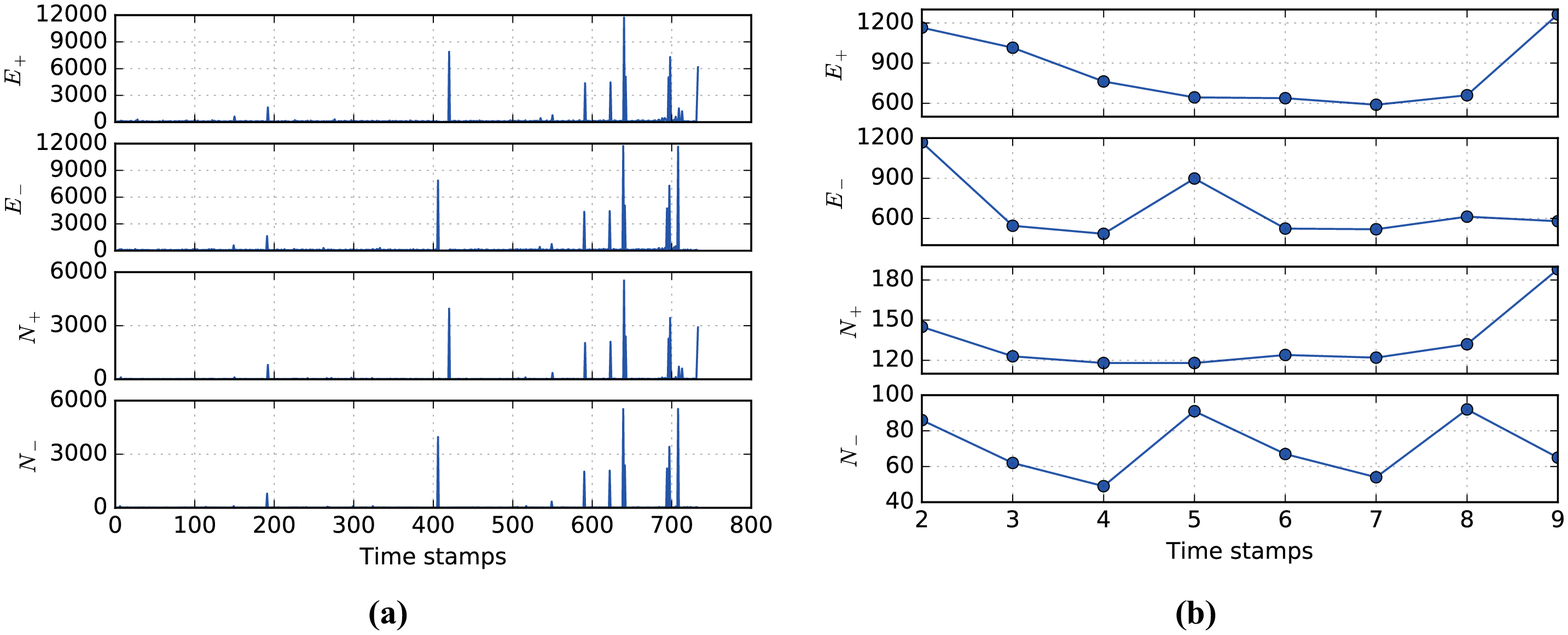}
	\caption{(Color online) {\bf Structural changes over time in the datasets of interest.} 
	(a) The structural changes over 733 snapshots in the AS-Internet dataset, including the number of edges added ($E_+$) and deleted ($E_-$), as well as the number of nodes involved in changes ($N_+,N_-$). (b) The structural changes in the AS-Oregon dataset over 9 snapshots.}
	\label{fig9}
\end{figure}

AS-Internet dataset is a communication network of who-talks-to-whom from the border gateway protocol logs of routers in the Internet. The dataset contains 733 daily instances of autonomous systems (AS) graph from November 8, 1997 to January 2, 2000. The largest graph (dataset from January 2, 2000) has 6474 nodes and 13859 edges. The nodes and edges are added or removed over time. Fig \ref{fig9} (a) shows the number of edges added and deleted, as well as the number of nodes involved in these changes. It is shown that the network topology can change dramatically at some snapshots.

AS-Oregon dataset contains nine undirected networks of AS peering information inferred from Oregon route-views between March 31, 2001 and May 26, 2001. These nine networks are different snapshots of the data with a minimum of 10,670 (March 31, 2001) and maximum of 11,174 (May 26, 2001) nodes. In addition, the number of edges ranges from 22,002 (April 7, 2001) to 23,409 (May 26, 2001). Fig \ref{fig9} (b) shows the number of edges added and deleted, as well as the number of nodes involved in the changes for the AS-Oregon dataset.

Since the real community structures of both datasets are unavailable, it is impossible to use NMI to evaluate the performance of different algorithms. Hence, we use \emph{modularity} \cite{PhysRevE.69.026113} to evaluate different algorithms on the datasets. In particular, we will show modularity values and processing times of ALPA in comparison with other methods. For each dataset, dynamic algorithms like ALPA (also iLCD) run on the network modifications, whereas the static method Infomap and snapshot method FacetNet have to be performed on the whole network snapshot at each time step.

\begin{figure}[!h]
	\centering
	\includegraphics[width=\linewidth]{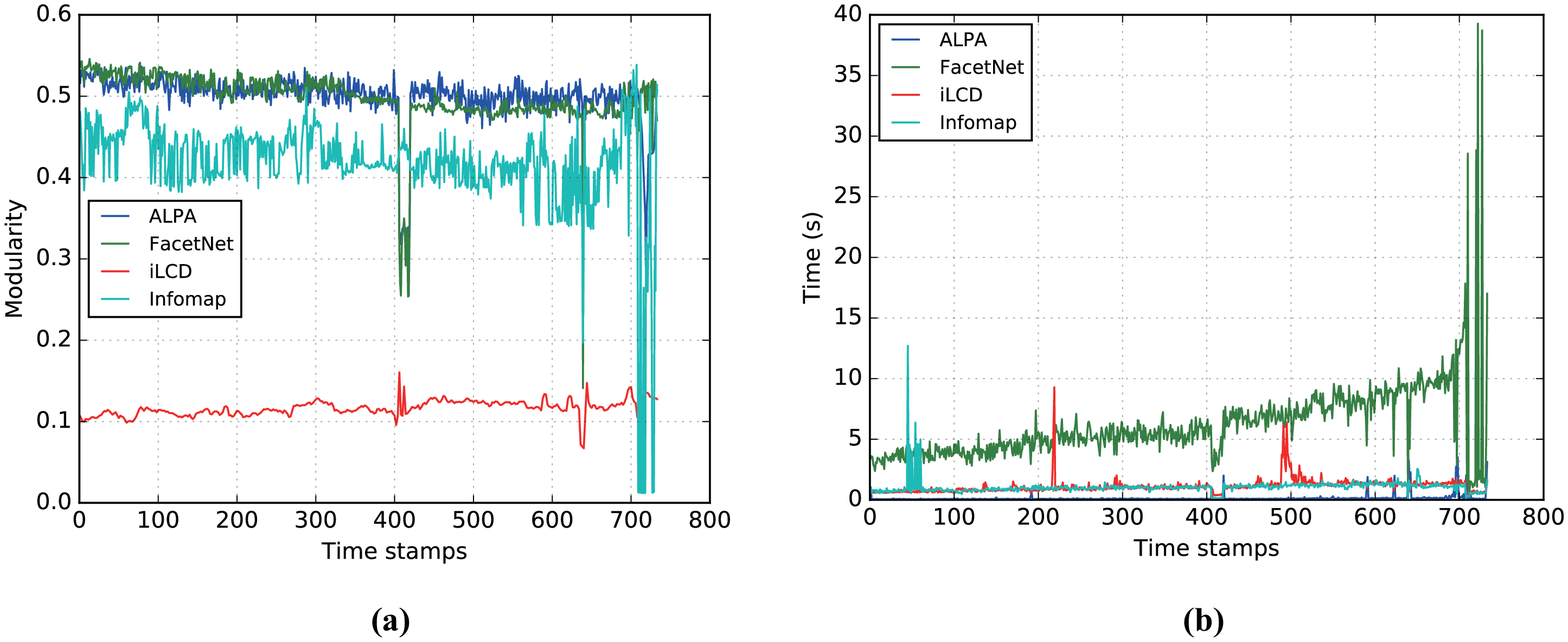}
	\caption{(Color online) {\bf Results on the AS-Internet dataset.} 
	Comparison of (a) modularity and (b) consuming time of ALPA at each snapshot with FacetNet, iLCD and Infomap on the AS-Internet dataset.}
	\label{fig10}
\end{figure}

Fig \ref{fig10} (a) shows the modularity values of ALPA and three other algorithms on the AS-Internet dataset. It is shown that ALPA and FacetNet have similar performance, and both of them achieve competitively higher modularity values than Infomap does for most of the snapshots. While iLCD fails to find strong community structure at all. In particular, the modularity values obtained by ALPA are more stable over time, since our method keeps preserving the community
structure of the previous snapshots and only considers current network changes. Retaining the historical information is a great advantage of ALPA because it avoids the expense of recomputing from scratch and makes the algorithm run faster. As shown in Fig \ref{fig10} (b), the computational cost is significantly reduced in ALPA. The running time of iLCD and Infomap is close. FacetNet requires a little more time. In fact, ALPA is three times faster than iLCD, two times faster than Infomap, and 250 times faster than FacetNet on the AS-Internet dataset. These results indicate that on the AS-Internet dataset, both ALPA and FacetNet are able to identify high quality community structure with high modularity. However, only our method significantly reduces the processing time.

\begin{figure}[!h]
	\centering
	\includegraphics[width=\linewidth]{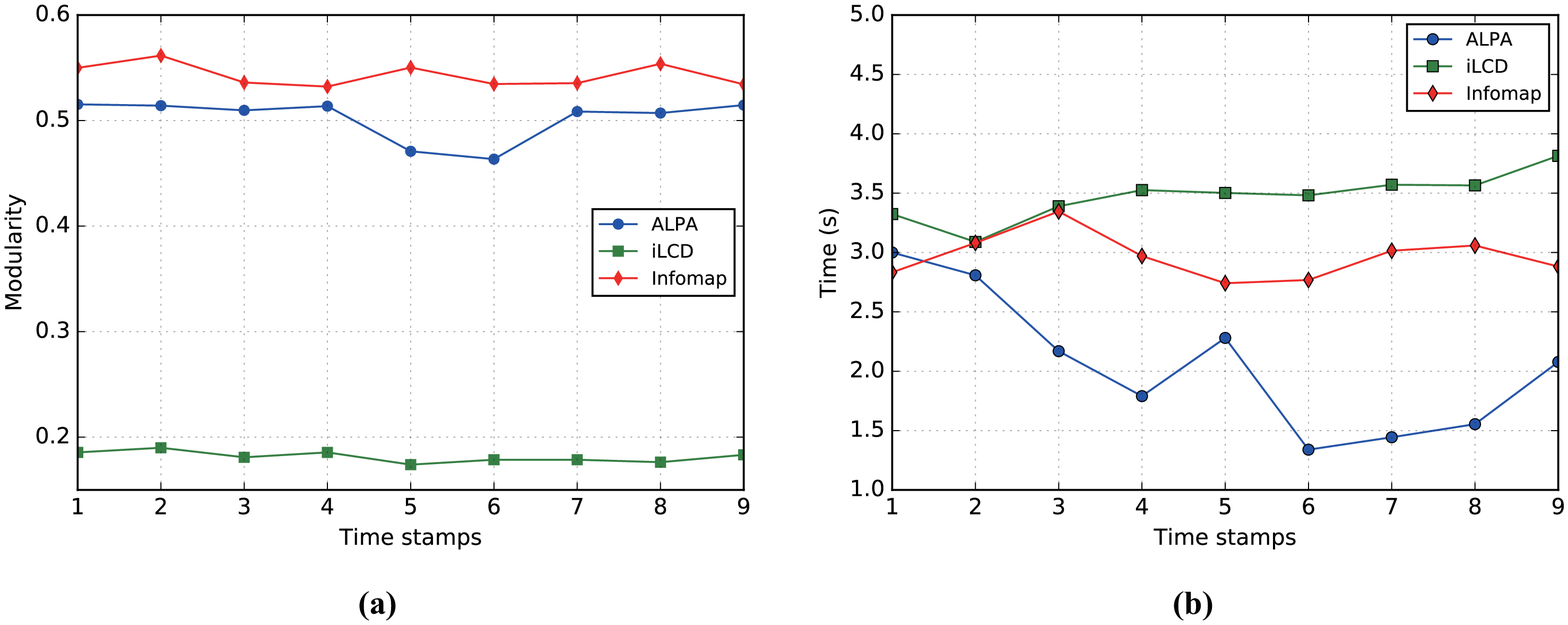}
	\caption{(Color online) {\bf Results on the AS-Oregon dataset.} 
	Comparison of modularity (a) and consuming time (b) of ALPA at each snapshot with iLCD and Infomap on the AS-Oregon dataset.}
	\label{fig11}
\end{figure}

Compared with AS-Internet dataset, AS-Oregon has fewer snapshots. However, the number of nodes and edges is large enough for an extensive analysis. In Fig \ref{fig11} (a), we compare modularity values obtained by ALPA at each network snapshot with those of iLCD and Infomap. FacetNet does not appear to complete the tasks due to the overflow in memory, and is thus excluded from the plots. It is shown that the modularity values obtained by ALPA are close to those
obtained by Infomap and are far higher than those obtained by iLCD. Fig \ref{fig11} (b) shows that regarding the running time ALPA outperforms Infomap as well as iLCD. In conclusion, high modularity values and low computational cost on this dataset confirm the effectiveness of our method.

\section*{Discussion}
In this work, we proposed an adaptive algorithm ALPA to detect communities in dynamic networks. It processes a sequence of modifications on the network and tries to maintain a fairly good community structure by updating a few existing communities through a local label propagation process, rather than computing the whole community decomposition from scratch. The advantages of our approach are as follows. Firstly, it requires neither any user-defined parameters nor the prior information of communities. Secondly, it is fast, scalable and incremental, i.e., it can work in a streaming fashion: whenever there is a modification of the network, ALPA adapts its result according to the modification by taking advantage of the historical information. Thirdly, it can monitor the evolution of each community at low computational cost. We have tested our method on synthetic networks and have shown that it identifies the planted communities with a high degree of success. We have also tested it on two real-world networks and have shown that it detects community structures with relatively high modularity scores and low computational costs.

It is difficult to accurately determine the time complexity of ALPA due to its randomness. We can roughly analyze the time complexity of ALPA as follows. The LLP or warm-up process is a local version of the original LPA. The time complexity is roughly $O(\langle m_c\rangle)$, where $\langle m_c\rangle$ is the mean number of edges in each community. Generally, $\langle m_c\rangle$ is tiny compared with the number of edges of the whole network. For each network modification, ALPA updates the current community structure with time complexity roughly $O(I\langle m_c\rangle)$, where $I$ is the number of iterations (usually is a small constant). The total time complexity of the ALPA for a dynamic network with $T$ time steps is $O(TI\langle m_c\rangle)$. Therefore, our method is applicable to analyzing communities of large dynamic networks which evolve rapidly, especially when sizes of communities are small.

As noticed, ALPA is not deterministic due to the random update order of nodes and a lot of tie-breaks. However, in our experiments, it is shown that ALPA is generally able to obtain the same partition in most runs. Only when community structure is not clear enough, may ALPA produce many similar partitions in multiple runs. We could obtain more robust and stable results by adopting a more deterministic update rule, e.g., considering similarities between adjacent node pairs when updating the labels of nodes.

In the forthcoming works, we plan to improve the current method by employing other more deterministic update rules and extend it to overlapping community detection. We also plan to apply our method to some practical applications, for example, constructing efficient distributed social-based message routing policy for wireless Ad hoc network.

\section*{Acknowledgments}
We thank Bo Gao, Guanglei Li for helpful comments on earlier drafts of the manuscript. This work was supported in partial by National Natural Science Foundation of China (Grant No. 11505071) and Program of Introducing Talents of Discipline to Universities (Grant No. B08033).


\nolinenumbers

%
%
%


\begin{thebibliography}{10}
	
	\bibitem{Strogatz2001}
	Strogatz SH.
	\newblock Exploring complex networks.
	\newblock Nature. 2001;410(6825):268--276.
	\newblock doi:{10.1038/35065725}.
	
	\bibitem{RevModPhys.74.47}
	Albert R, Barab\'asi AL.
	\newblock Statistical mechanics of complex networks.
	\newblock Rev Mod Phys. 2002;74:47--97.
	\newblock doi:{10.1103/RevModPhys.74.47}.
	
	\bibitem{doi:10.1080/00018730110112519}
	Dorogovtsev SN, Mendes JFF.
	\newblock Evolution of networks.
	\newblock Advances in Physics. 2002;51(4):1079--1187.
	\newblock doi:{10.1080/00018730110112519}.
	
	\bibitem{doi:10.1137/S003614450342480}
	Newman MEJ.
	\newblock The Structure and Function of Complex Networks.
	\newblock SIAM Review. 2003;45(2):167--256.
	\newblock doi:{10.1137/S003614450342480}.
	
	\bibitem{Faloutsos:1999:PRI:316194.316229}
	Faloutsos M, Faloutsos P, Faloutsos C.
	\newblock On Power-law Relationships of the Internet Topology.
	\newblock SIGCOMM Comput Commun Rev. 1999;29(4):251--262.
	\newblock doi:{10.1145/316194.316229}.
	
	\bibitem{Amaral10102000}
	Amaral LAN, Scala A, Barthélémy M, Stanley HE.
	\newblock Classes of small-world networks.
	\newblock Proceedings of the National Academy of Sciences.
	2000;97(21):11149--11152.
	\newblock doi:{10.1073/pnas.200327197}.
	
	\bibitem{Newman16012001}
	Newman MEJ.
	\newblock The structure of scientific collaboration networks.
	\newblock Proceedings of the National Academy of Sciences. 2001;98(2):404--409.
	\newblock doi:{10.1073/pnas.98.2.404}.
	
	\bibitem{Williams2000}
	Williams RJ, Martinez ND.
	\newblock Simple rules yield complex food webs.
	\newblock Nature. 2000;404(6774):180--183.
	\newblock doi:{10.1038/35004572}.
	
	\bibitem{Dunne01102002}
	Dunne JA, Williams RJ, Martinez ND.
	\newblock Food-web structure and network theory: The role of connectance and
	size.
	\newblock Proceedings of the National Academy of Sciences.
	2002;99(20):12917--12922.
	\newblock doi:{10.1073/pnas.192407699}.
	
	\bibitem{Jeong2000}
	Jeong H, Tombor B, Albert R, Oltvai ZN, Barabasi AL.
	\newblock The large-scale organization of metabolic networks.
	\newblock Nature. 2000;407(6804):651--654.
	\newblock doi:{10.1038/35036627}.
	
	\bibitem{Fell2000}
	Fell DA, Wagner A.
	\newblock The small world of metabolism.
	\newblock Nat Biotech. 2000;18(11):1121--1122.
	\newblock doi:{10.1038/81025}.
	
	\bibitem{Girvan11062002}
	Girvan M, Newman MEJ.
	\newblock Community structure in social and biological networks.
	\newblock Proceedings of the National Academy of Sciences.
	2002;99(12):7821--7826.
	\newblock doi:{10.1073/pnas.122653799}.
	
	\bibitem{1742-5468-2005-09-P09008}
	Danon L, Díaz-Guilera A, Duch J, Arenas A.
	\newblock Comparing community structure identification.
	\newblock Journal of Statistical Mechanics: Theory and Experiment.
	2005;2005(09):P09008.
	
	\bibitem{Fortunato201075}
	Fortunato S.
	\newblock Community detection in graphs.
	\newblock Physics Reports. 2010;486(3–5):75 -- 174.
	\newblock doi:{http://dx.doi.org/10.1016/j.physrep.2009.11.002}.
	
	\bibitem{Newman06062006}
	Newman MEJ.
	\newblock Modularity and community structure in networks.
	\newblock Proceedings of the National Academy of Sciences.
	2006;103(23):8577--8582.
	\newblock doi:{10.1073/pnas.0601602103}.
	
	\bibitem{FastUnfolding2008}
	Blondel VD, Guillaume JL, Lambiotte R, Lefebvre E.
	\newblock Fast unfolding of communities in large networks.
	\newblock Journal of Statistical Mechanics: Theory and Experiment.
	2008;2008(10):P10008.
	
	\bibitem{PhysRevE.76.036106}
	Raghavan UN, Albert R, Kumara S.
	\newblock Near linear time algorithm to detect community structures in
	large-scale networks.
	\newblock Phys Rev E. 2007;76:036106.
	\newblock doi:{10.1103/PhysRevE.76.036106}.
	
	\bibitem{1367-2630-12-10-103018}
	Gregory S.
	\newblock Finding overlapping communities in networks by label propagation.
	\newblock New Journal of Physics. 2010;12(10):103018.
	
	\bibitem{JieruiXie-LabelRank-NSW:2013}
	Xie J, Szymanski BK.
	\newblock LabelRank: A stabilized label propagation algorithm for community
	detection in networks.
	\newblock In: Network Science Workshop (NSW), 2013 IEEE 2nd; 2013. p. 138--143.
	
	\bibitem{PhysRevE.83.016107}
	Karrer B, Newman MEJ.
	\newblock Stochastic blockmodels and community structure in networks.
	\newblock Phys Rev E. 2011;83:016107.
	\newblock doi:{10.1103/PhysRevE.83.016107}.
	
	\bibitem{Aldecoa2013surprise}
	Aldecoa R, Mar{\'i}n I.
	\newblock Surprise maximization reveals the community structure of complex
	networks.
	\newblock Scientific Reports. 2013;3:1060.
	
	\bibitem{10.1371/journal.pone.0018961}
	Lancichinetti A, Radicchi F, Ramasco JJ, Fortunato S.
	\newblock Finding Statistically Significant Communities in Networks.
	\newblock PLoS ONE. 2011;6(4):1--18.
	\newblock doi:{10.1371/journal.pone.0018961}.
	
	\bibitem{Singh2015}
	Singh A, Humphries MD.
	\newblock Finding communities in sparse networks.
	\newblock Scientific Reports. 2015;5:8828.
	
	\bibitem{Rosvall01052007}
	Rosvall M, Bergstrom CT.
	\newblock An information-theoretic framework for resolving community structure
	in complex networks.
	\newblock Proceedings of the National Academy of Sciences.
	2007;104(18):7327--7331.
	\newblock doi:{10.1073/pnas.0611034104}.
	
	\bibitem{Rosvall29012008}
	Rosvall M, Bergstrom CT.
	\newblock Maps of random walks on complex networks reveal community structure.
	\newblock Proceedings of the National Academy of Sciences.
	2008;105(4):1118--1123.
	\newblock doi:{10.1073/pnas.0706851105}.
	
	\bibitem{Zalik2015}
	\v{Z}alik KR.
	\newblock Maximal Neighbor Similarity Reveals Real Communities in Networks.
	\newblock Scientific Reports. 2015;5:18374.
	
	\bibitem{PhysRevE.72.046108}
	Bagrow JP, Bollt EM.
	\newblock Local method for detecting communities.
	\newblock Phys Rev E. 2005;72:046108.
	\newblock doi:{10.1103/PhysRevE.72.046108}.
	
	\bibitem{5562773}
	Greene D, Doyle D, Cunningham P.
	\newblock Tracking the Evolution of Communities in Dynamic Social Networks.
	\newblock In: Advances in Social Networks Analysis and Mining (ASONAM), 2010
	International Conference on; 2010. p. 176--183.
	
	\bibitem{Asur:2007:EFC:1281192.1281290}
	Asur S, Parthasarathy S, Ucar D.
	\newblock An Event-based Framework for Characterizing the Evolutionary Behavior
	of Interaction Graphs.
	\newblock In: Proceedings of the 13th ACM SIGKDD International Conference on
	Knowledge Discovery and Data Mining. KDD '07. New York, NY, USA: ACM; 2007.
	p. 913--921.
	\newblock Available from: \url{http://doi.acm.org/10.1145/1281192.1281290}.
	
	\bibitem{ning2007incremental}
	Ning H, Xu W, Chi Y, Gong Y, Huang T.
	\newblock Incremental spectral clustering with application to monitoring of
	evolving blog communities.
	\newblock In: Proceedings of the 2007 SIAM International Conference on Data
	Mining. SIAM; 2007. p. 261--272.
	
	\bibitem{Palla2007}
	Palla G, Barabasi AL, Vicsek T.
	\newblock Quantifying social group evolution.
	\newblock Nature. 2007;446(7136):664--667.
	\newblock doi:{10.1038/nature05670}.
	
	\bibitem{6758385}
	Xu KS, Hero AO.
	\newblock Dynamic Stochastic Blockmodels for Time-Evolving Social Networks.
	\newblock IEEE Journal of Selected Topics in Signal Processing.
	2014;8(4):552--562.
	\newblock doi:{10.1109/JSTSP.2014.2310294}.
	
	\bibitem{Lin:2009:ACE:1514888.1514891}
	Lin YR, Chi Y, Zhu S, Sundaram H, Tseng BL.
	\newblock Analyzing Communities and Their Evolutions in Dynamic Social
	Networks.
	\newblock ACM Trans Knowl Discov Data. 2009;3(2):8:1--8:31.
	\newblock doi:{10.1145/1514888.1514891}.
	
	\bibitem{yang2009bayesian}
	Yang T, Chi Y, Zhu S, Gong Y, Jin R.
	\newblock A Bayesian Approach Toward Finding Communities and Their Evolutions
	in Dynamic Social Networks.
	\newblock In: SDM. vol.~9; 2009. p. 990--1001.
	
	\bibitem{Sun:2007:GPM:1281192.1281266}
	Sun J, Faloutsos C, Papadimitriou S, Yu PS.
	\newblock GraphScope: Parameter-free Mining of Large Time-evolving Graphs.
	\newblock In: Proceedings of the 13th ACM SIGKDD International Conference on
	Knowledge Discovery and Data Mining. KDD '07. New York, NY, USA: ACM; 2007.
	p. 687--696.
	\newblock Available from: \url{http://doi.acm.org/10.1145/1281192.1281266}.
	
	\bibitem{iLCD}
	Cazabet R, Amblard F, Hanachi C.
	\newblock Detection of Overlapping Communities in Dynamical Social Networks.
	\newblock In: Social Computing (SocialCom), 2010 IEEE Second International
	Conference on; 2010. p. 309--314.
	
	\bibitem{10.1371/journal.pone.0091431}
	Nguyen NP, Dinh TN, Shen Y, Thai MT.
	\newblock Dynamic Social Community Detection and Its Applications.
	\newblock PLoS ONE. 2014;9(4):1--18.
	\newblock doi:{10.1371/journal.pone.0091431}.
	
	\bibitem{10.1371/journal.pone.0086891}
	Lung RI, Chira C, Andreica A.
	\newblock Game Theory and Extremal Optimization for Community Detection in
	Complex Dynamic Networks.
	\newblock PLoS ONE. 2014;9(2):1--11.
	\newblock doi:{10.1371/journal.pone.0086891}.
	
	\bibitem{xie2013labelrankt}
	Xie J, Chen M, Szymanski BK.
	\newblock LabelrankT: Incremental community detection in dynamic networks via
	label propagation.
	\newblock In: Proceedings of the Workshop on Dynamic Networks Management and
	Mining. ACM; 2013. p. 25--32.
	
	\bibitem{6004645}
	Xie J, Szymanski BK.
	\newblock Community detection using a neighborhood strength driven Label
	Propagation Algorithm.
	\newblock In: Network Science Workshop (NSW), 2011 IEEE; 2011. p. 188--195.
	
	\bibitem{Leskovec:2005:GOT:1081870.1081893}
	Leskovec J, Kleinberg J, Faloutsos C.
	\newblock Graphs over Time: Densification Laws, Shrinking Diameters and
	Possible Explanations.
	\newblock In: Proceedings of the Eleventh ACM SIGKDD International Conference
	on Knowledge Discovery in Data Mining. KDD '05. New York, NY, USA: ACM; 2005.
	p. 177--187.
	\newblock Available from: \url{http://doi.acm.org/10.1145/1081870.1081893}.
	
	\bibitem{PhysRevE.78.046110}
	Lancichinetti A, Fortunato S, Radicchi F.
	\newblock Benchmark graphs for testing community detection algorithms.
	\newblock Phys Rev E. 2008;78:046110.
	\newblock doi:{10.1103/PhysRevE.78.046110}.
	
	\bibitem{10.1371/journal.pone.0137502}
	Mall R, Langone R, Suykens JAK.
	\newblock Netgram: Visualizing Communities in Evolving Networks.
	\newblock PLoS ONE. 2015;10(9):1--24.
	\newblock doi:{10.1371/journal.pone.0137502}.
	
	\bibitem{PhysRevE.92.012805}
	Granell C, Darst RK, Arenas A, Fortunato S, G\'omez S.
	\newblock Benchmark model to assess community structure in evolving networks.
	\newblock Phys Rev E. 2015;92:012805.
	\newblock doi:{10.1103/PhysRevE.92.012805}.
	
	\bibitem{snapnets}
	Leskovec J, Krevl A. {SNAP Datasets}: {Stanford} Large Network Dataset
	Collection; 2014.
	\newblock \url{http://snap.stanford.edu/data}.
	
	\bibitem{PhysRevE.69.026113}
	Newman MEJ, Girvan M.
	\newblock Finding and evaluating community structure in networks.
	\newblock Phys Rev E. 2004;69:026113.
	\newblock doi:{10.1103/PhysRevE.69.026113}.
	
\end{thebibliography}

\end{document}